# Self-consistent theory of nanodomain formation on non-polar surfaces of ferroelectrics

**Anna N. Morozovska[1][*], Anton Ievlev[2], Vyacheslav V. Obukhovskii[3], Yevhen Fomichov[4], Oleksandr V. Varenyk[1], Vladimir Ya. Shur[5], Sergei V. Kalinin[2] and Eugene A. Eliseev[6][†]**

[1] Institute of Physics, National Academy of Science of Ukraine,
46, pr. Nauky, 03028 Kyiv, Ukraine

[3]Taras Shevchenko Kyiv National University, Radiophysical Faculty, 4g, pr. Akademika Hlushkova, 03022 Kyiv, Ukraine

[2] The Center for Nanophase Materials Sciences, Oak Ridge National Laboratory,
Oak Ridge, TN 37831

[3] Taras Shevchenko Kiev National University, Radiophysical Faculty
4g, pr. Akademika Hlushkova, 03022 Kiev, Ukraine

[4] National Technical University of Ukraine "Kyiv Polytechnic Institute", Faculty of Electronics, 16 st. Politekhnichna, block number 12, 03056, Kyiv, Ukraine

[6] Institute for Problems of Materials Science, National Academy of Science of Ukraine,
3, Krjijanovskogo, 03142 Kyiv, Ukraine

## Abstract

We propose a self-consistent theoretical approach capable to describe the peculiarities of the anisotropic nanodomain formation induced by a charged AFM probe on non-polar cuts of ferroelectrics. The proposed semi-phenomenological approach accounts for the difference of the threshold fields required for the domain wall motion along non-polar X- and Y – cuts, and polar Z-cut of $LiNbO_3$. The effect steams from the fact, that the minimal distance between the equilibrium atomic positions of domain wall and the profile of lattice pinning barrier appeared different for different directions due to the crystallographic anisotropy.

Using relaxation-type equation with cubic nonlinearity we calculated the polarization reversal dynamics during the probe-induced nanodomain formation for different threshold field values. The different velocity of domain growth and consequently equilibrium domain sizes on X-, Y- and Z-cuts of $LiNbO_3$ originate from the anisotropy of the threshold field. Note that the smaller is the threshold field the larger are the domain sizes, and the fact allows explaining several times difference in nanodomain length experimentally observed on X- and Y-cuts of $LiNbO_3$. Obtained results can give insight into the nanoscale anisotropic dynamics of polarization reversal in strongly inhomogeneous electric field.

---

[*] Corresponding author 1: anna.n.morozovska@gmail.com

[†] Corresponding author 2: eugene.a.eliseev@gmail.com

**I. Introduction**

The investigation of local polarization dynamics in ferroelectric materials becomes one of the most intriguing and rapidly developed direction of fundamental studies in nano-physics as well as prospective for next generation of memory devices [1, 2, 3, 4, 5, 6]. The reason that made the investigations very attractive is the possibility to control the local redistribution of ferroelectric polarization, in particular to form the nanodomains arrays by the scanning probe atomic force microscopy (AFM) [7, 8]. Actually, the strongly inhomogeneous electric field of the charged AFM probe is the most appropriate source of the nanodomain formation [2].

There are many experimental and theoretical studies of nanodomain formation on polar surfaces of ferroelectric crystals by the biased AFM probe, demonstrating that the normal [9, 10, 11, 12, 13, 14, 15] or anomalous [16, 17, 18] local polarization reversal can take place along polar axes. Experimental studies of the micro- and nano- domain walls motion have been performed in typical crystalline ferroelectric materials such as Pb(Zr, Ti)$O_3$, $Pb_5Ge_3O_{11}$, $LiTaO_3$, $LiNbO_3$ [19, 20, 21, 22]. Appeared that the micro- and nano-domain lateral size is linearly proportional to the voltage pulse amplitude and is to the logarithm of the pulse duration [7, 9].

A sizable amount of semi-phenomenological models of the nanodomain formation caused by the inhomogeneous electric field of AFM probe was proposed. Mainly these models can be divided in two different groups, namely Landauer-Molotskii (LM) energetic approach [8, 23, 24, 25, 26, 27, 28] and Landau-Ginzburg-Devonshire (LGD) approach [29, 30, 31, 32]. In order to obtain analytical expressions for the free energy, LM approach considers the semi-ellipsoidal domain with infinitely thin walls and includes the domain wall surface energy into the free energy functional. Besides the domain wall surface energy that is simply proportional to the domain surface, the functional includes analytical expressions for the electrostatic depolarization field energy and the interaction energy of the domain polarization with a charged probe electric field. Free energy minimization gives transcendental equations for the ellipsoid semi-axes regarded as domain sizes.

LGD approach allows calculating the domain shape, sizes and the wall thicknesses as well as the electric field distribution in the system in a self-consistent way, as a solution of the relaxation-type non-linear time-dependent differential Landau-Ginzburg-Devonshire (TD-LGD) equation for the temporal evolution of ferroelectric polarization spatial distribution, coupled with the Poisson equation for the electric field, bound and space charges. Combined with powerful phase-field method, TD-LGD approach allows obtaining rigorous numerical results for domain kinetics and approximate analytics in the sense of interpolation [5]. LGD-approach considers the nanodomain formation process in a self-consistent manner accounting for the fact that the width of the growing domain wall is determined by its bound charges, which distribution depends on the angles near the curved apexes of the nanodomains. The corrections originated from the finite width effect appeared far not small and exactly leads to the intrinsic domain breakdown effect [30]. In particular, numerical calculations of the electric field near the charged domain wall explains the domain growth in the areas with external electric field well below coercive one [30, 32] and so confirms the domain breakdown effect on sub-micron and micron distances observed experimentally in LiNbO$_3$ for different geometry [8].

Let us underline that almost all available experimental and theoretical works are devoted to the investigation of the nanodomain kinetics on polar surfaces of ferroelectric crystals; at the same time the forward growth remains one of the most unexplored stages due to lack of experimental methods allowing to study it. Only recently Ievlev et al [32] and Alikin et al [33] demonstrated that the probe-induced polarization reversal on X- and Y- nonpolar cuts in single crystal of congruent LiNbO$_3$ can give insight in the forward growth with high spatial resolution. They reported about the strong deviation of the domain shape from a semi-ellipsoid as well as the difference of the domain shape and length on X- and Y-cut (see **Figure 1**), which contradicts to the recent theoretical estimations of Pertsev et al [34], performed in the framework of thermodynamic LM model. Alikin et al concluded that their results can be explained only in terms of kinetic approach, which self-consistent formulation was absent.

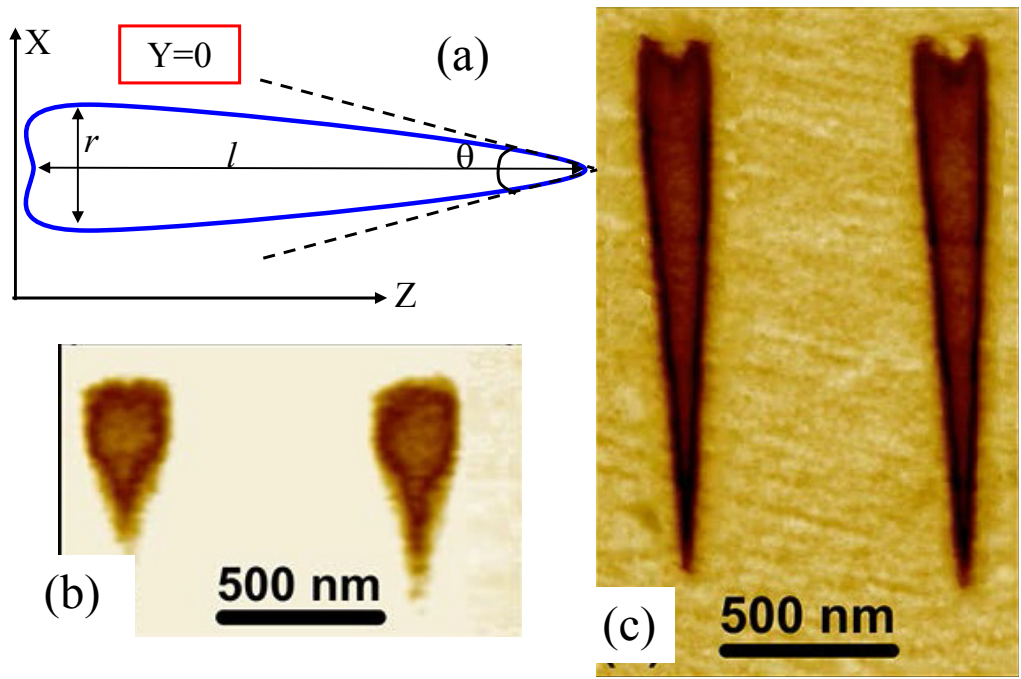

**Figure 1**. **(a)** Sketch of the domain shape in the XZ plane (Y=0) induced by AFM probe on congruent LiNbO$_3$ (CLN) non-polar cuts. **(b-c)** Experimental results from Alikin et al [33] showing domains formed as a result of tip-induced switching by single rectangular pulses with amplitude $U_{sw}$ = 80 V and duration $t_{sw}$ = 1 s on **(b)** X- and **(c)** Y-cuts of 20-μm-thick CLN [33]. *[permission for b-c will be granted]*

It is well-established that the domain wall kinetics is strongly affected by the lattice pinning [35], which was not accounted in all theoretical studies devoted to the nanodomain formation at non-polar cuts. Lattice pinning phenomenon consists in the fact that the domain wall can move over a distance which is a multiple of the lattice constant. The critical electric field, that should be applied for the local polarization reversal in the nanoscale, is rather defined by the interplay between the pinning, depolarization and probe electric field that the intrinsic thermodynamic field [31]. Conventionally, the critical fields can be estimated analytically using several approaches. Suzuki-Ishibashi (S-I) model [36] can be used for the threshold field determination. The activation filed, that determines the nucleation process kinetics, can be defined within Miller-Weinreich [37] or Burtsev-Chervonobrodov (B-C) approach [38] modified by Rappe et al [39] and Aravind et al [31] allowing for its dependence on the polarization gradient and depolarization effects at the wall.

The necessity of constructing a theory that adequately describes the formation of nanodomains on non-polar surfaces is dictated not only by the general scientific interest, but is of great practical importance for determining the domain structure on the surface and for the control of local polarization dynamics in the nanoscale by AFM. The goal motivates us to perform the study self-consistent modeling of the nanodomain formation on the non-polar surfaces of LiNbO$_3$ with a special attention to the lattice pinning anisotropy.

The manuscript is organized in the following way. The introductive section I is followed by the section II, which contains three subsections devoted to the description of the problem statement for self-consistent numerical modeling, atomistic toy models explaining the threshold field anisotropy from the lattice pinning anisotropy and theoretical background for interpolation functions used for the domain sizes description. Results of numerical modeling of polarization dynamics is presented in the section III. Section IV demonstrates the capability of our approach to describe quantitatively experimental results [33]. Section V is a brief summary.

## II. Theoretical description
### 2.1. Problem statement for self-consistent numerical modeling

Schematics of the probe-induced nanodomain reversal on the non-polar cuts surface of LiNbO$_3$ is shown in the **Figure 2**. The radial component of the probe electric field induces the domain nucleation and growth for the considered geometry. For an axially-symmetric probe tip the radial component of its electric field $E_z$ is anti-symmetric, and the field maximum is located at some distance from the probe axes, as schematically shown in the figure.

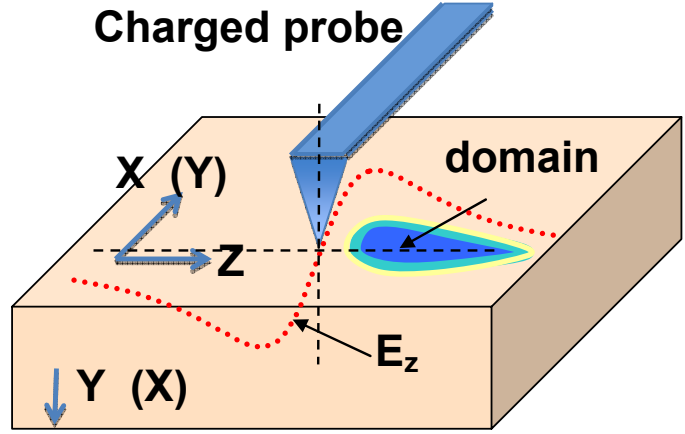

**Figure 2.** Schematics of the probe-induced nanodomain switching near the non-polar cuts surface of LiNbO$_3$. *[adapted from [30]]*

The potential $\varphi$ of quasi-stationary electric field, $\mathbf{E} = -\nabla\varphi$, $\varphi$ satisfies electrostatic equations inside the layered system. Outside the probe, in the air/vacuum ambient semispace, $-\infty < y < 0$, $\varphi$ satisfies Laplace equation

$$\left(\frac{\partial^2}{\partial z^2} + \frac{\partial^2}{\partial y^2} + \frac{\partial^2}{\partial x^2}\right)\varphi_d = 0, \qquad (1)$$

The equipotential surface corresponds to the biased conducting probe surface, $\varphi_d|_{tip} = U$ and U is the periodic voltage applied to the probe, $U(t) = V\sin(\omega t)$. The potential satisfies an anisotropic Poisson equation inside a ferroelectric layer $0 < y < L$:

$$\varepsilon_{33}^b \frac{\partial^2 \varphi_f}{\partial z^2} + \varepsilon_{11}^f \left(\frac{\partial^2 \varphi_f}{\partial x^2} + \frac{\partial^2 \varphi_f}{\partial y^2}\right) = \frac{1}{\varepsilon_0}\left(\frac{\partial P_z}{\partial z} - \rho\right), \qquad (2)$$

$\varepsilon_{33}^b$ is the background dielectric permittivity of the ferroelectric, that is considered hereinafter without free carriers ($\rho$=0). The gradient term $\partial P_z/\partial z$ reflects the existence of the bound charges originated from the inhomogeneous ferroelectric polarization $P_z(x,y,z)$.

Equations (1) and (2) should be supplemented with the boundary conditions of zero potential at bottom planar electrode, $\varphi_f(x, y = L, z) = 0$, continuous potential on the interface between air and ferroelectric, $\varphi_f(x, y = 0, z) = \varphi_d(x, y = 0, z)$ and normal displacement, $\left. D_y^d - D_y^f \right|_{y=0} = 0$, where the displacement components $D_y^f = -\varepsilon_0 \varepsilon_{11}^f (\partial \varphi_f / \partial y)$ and $D_y^d = -\varepsilon_0 (\partial \varphi_d / \partial y)$.

We regard that the ferroelectric polarization dynamics obeys relaxation-type differential equation with cubic nonlinearity [32]:

$$\tau_0 \frac{\partial \tilde{P}_z}{\partial t} - \tilde{P}_z + \tilde{P}_z^3 - R_c^2\left(\frac{\partial^2 \tilde{P}_z}{\partial z^2} + \frac{\partial^2 \tilde{P}_z}{\partial y^2} + \frac{\partial^2 \tilde{P}_z}{\partial x^2}\right) = \tilde{E}_z \qquad (3)$$

Here $\tilde{P}_z = P_z/P_S$ is the normalized ferroelectric polarization directed along z-axes, normalized on the spontaneous polarization $P_S$. Characteristic time $\tau_0 = -\Gamma/\alpha$ is determined by the ratio of kinetic Khalatnikov coefficient $\Gamma$ and generalized dielectric stiffness $\alpha$. In accordance with the Curie-Weiss law, the coefficient $\alpha = \alpha_T(T - T_C)$, where T is temperature in Kelvins and $T_C$ is Curie temperature. The time $\tau_0$ is in fact a soft phonon time that is small enough far from the Curie temperature (e.g. at room for LiNbO$_3$). Correlation length $R_c = \sqrt{-g/\alpha}$ is about 1 nm well below the ferroelectric phase transition temperature, where g is the gradient coefficient in the LGD potential.

Equation (3) should be supplemented with the boundary conditions corresponding to the uniform polarization far from the probe apex, $\tilde{P}_z(r \to \infty) = +1$ (and so $\partial \tilde{P}_z/\partial x_i \to 0$), and natural boundary conditions at the ferroelectric surfaces, $\partial P_z/\partial y\big|_{y=0} = 0$ and $\partial P_z/\partial y\big|_{y=L} = 0$.

The right-hand-side of the equation (3) contains the electric field normalized on the "threshold" field $E_{th}$ of the domain wall motion, $\tilde{E}_z = E_z/E_{th}$. The uncharged domain wall is unpinned by lattice defects above the threshold field. In the case of domain nucleation the wall is inevitably charged at least in the vicinity of the growing domain apex. Hence the field $E_z$ is the sum of external probe field, screening and depolarization ones, and these three contributions in total is equal to $-\partial \varphi/\partial x_3$.

Note that despite the mathematical form of Equation (3) is the same as the form of Landau-Ginzburg-Devonshire-Khalatnikov equation, in fact we use a principal physical difference between them here and previously [32]. Namely, the thermodynamic coercive field is determined by the LGD-potential expansion coefficients $\alpha$ and $\beta$ as $E_c = 2\sqrt{-\alpha^3/27\beta}$. Thus it is independent on the domain wall growth direction [30] and mostly gives strongly overestimated values in comparison with experimentally observed ones. In contrast to the LGD approach, hereinafter we suppose that the threshold field $E_{th}$ is determined by the lattice pinning (or Peierls barrier) that should depend on the domain wall type and orientation with respect to the crystallographic axes, as well as on the minimal distance between the equilibrium atomic positions of uncharged domain wall plane. The field $E_{th}$ can be much smaller than the thermodynamic coercive field, and it can be successfully fitted to experiment.

## 2.2. Atomistic toy model explaining the threshold field anisotropy

The main idea of our research is to use the fact, that the threshold field can be essentially different for the LiNbO$_3$ crystallographic X-cut (ZY plane), Y-cut (ZX plane) and Z-cut (XY plane), because the inter-atomic relief and energy barriers are anisotropic. 3D-atomic structure of LiNbO$_3$ crystallographic cuts are reconstructed in the **Figure 3a-c** using the coordinates from Boysen and Altorfer [40]. A suggested step-like path of the domain wall motion in the polar direction Z on the non-polar X- and Y-cuts is shown by an elementary step in the **Figures 3b-3c.** The separate schematic is shown in the **Figure 3d.** The suggestion about the step-like path is in an agreement with the schematics proposed by Alikin et al (see figure 4 in [33]). Using the path one can define the elementary step as Li-Li distances in the directions perpendicular to the polar Z-axes. Namely, using LiNbO$_3$ rhombohedra lattice parameters ($a$ = 5.15 Å, $c$ = 13.86 Å, angle $\alpha_c$ = 55°53' [41, 42]), XY-, XZ- and XY-planes schematics one can calculate the minimal distances between the equilibrium positions of uncharged domain wall planes at different crystallographic cuts, hereinafter denoted as $p_{[abc]}$. Namely, $p_{[100]} = \sqrt{3}a/2 \approx 4.460$ Å for X-cut, $p_{[010]} = a/2 \approx 2.575$ Å for Y-cut and $p_{[001]} = c/6 \approx 2.310$ Å for Z-cut. Hereinafter we associate [001] – Z-cut, [010] – Y-cut, [001] – X-cut.

In accordance with DFT calculations corresponding equilibrium position of the uncharged domain wall is determined as the center between two anion planes [31]. Calculated energy relief of Y walls in LiNbO$_3$ is schematically shown in the **Figure 3e**.

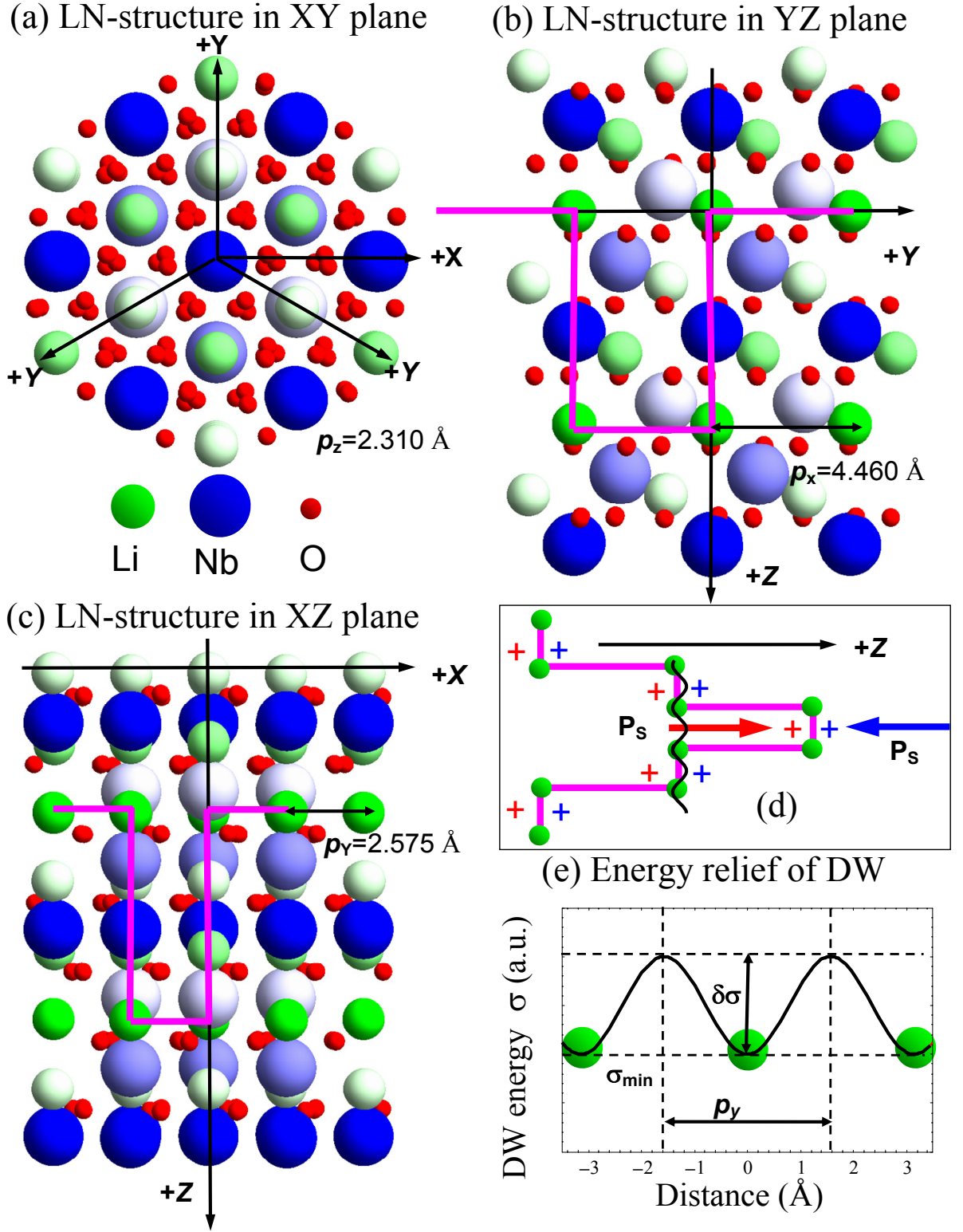

**Figure 3**. Atomic structure of LiNbO$_3$ Z-cut (XY plane) **(a)** X-cut (ZY plane) **(b)** and Y-cut (ZX-plane) **(c)**. **(d)** A suggested step-like path of the domain wall motion in the polar direction Z at the non-polar X- and Y-cuts. **(e)** Energy relief of Y-walls in LiNbO$_3$ *[adapted from [31]]*.

In the framework of S-I approach [35] the threshold field acquires the form

$$E_{th}^{[abc]} = -e^4 (\pi/2)^{7/2} \alpha P_S \left(w/p_{[abc]}\right)^3 \exp\left(-\pi^2 w/p_{[abc]}\right), \qquad (4)$$

Here the half-width of the domain wall $w$ is normalized on the minimal distance $p_{[abc]}$ between the equilibrium positions of the uncharged domain wall plane propagating in the crystallographic direction [abc]. The threshold field was calculated within Suzuki-Ishibashi model for LiNbO$_3$ parameters $\alpha$, $P_S$ and different domain wall half-width $w$, since the latter can be strongly affected by depolarization field and depends on the wall bound charge (e.g. incline angle with respect to the polar direction). Results are shown in the **Figure 4a**. As one can see, the value of $E_{th}$ differs on the one or even several orders of magnitude for different direction of the domain wall motion. Also the threshold field strongly decreases with $p_{[abc]}$ increase and vary in the range ($10^{-3}$ – $10^{+2}$) kV/mm. The threshold field $E_{th}$ monotonically and rapidly decreases with the wall half-width $w$ increase at $w > 1$ Å for any period $p_{[abc]}$. Note, that smaller $w$ values are unlikely physical. At fixed $w > 1$ Å the highest fields correspond to the smallest period $p_{[abc]}$, i.e. $E_{th}^Z < E_{th}^Y \ll E_{th}^X$ since $p_Z < p_Y < p_X$. This exactly means that the threshold field is the smallest for Z-cut, intermediate for Y-cut and the highest for the X-cut of the crystal.

The activation field, that determines the nucleation process kinetics, can be estimated within modified B-C approach as [31]:

$$E_a^{[abc]} = \frac{1}{V_0} \frac{\gamma R}{U P_S} \sqrt{\ln\left(\frac{\gamma R \sqrt{\sigma_{\min} \delta\sigma}}{2 p_{[abc]} P_S U}\right) \frac{\left(\sqrt{\sigma_{\min} \delta\sigma} p_{[abc]}\right)^3}{4\pi\varepsilon_0 \varepsilon_{11}}}. \quad (5)$$

Here $V_0$ is the elementary volume, $U$ is the voltage applied to the probe, $R$ is the effective probe size, $\gamma$ is the dielectric anisotropy factor; $\sigma_{\min}$ is the minimal value of the periodic lattice potential and $\delta\sigma$ is the modulation depth of the domain wall energy $\sigma_W(x) \approx \sigma_{\min} + \delta\sigma \sin^2(\pi(x - x_0)/p_{[abc]})$ (see **Figure 3e**). The critical activation voltage of domain nucleation, $U_{cr}$, can be defined from the requirement of $\ln\left(\frac{\gamma R \sqrt{\sigma_{\min} \delta\sigma}}{2 p_{[abc]} P_S U}\right) \geq 0$ in Eq.(5) that gives

$$U_{cr} = \frac{\gamma R \sqrt{\sigma_{\min} \delta\sigma}}{2 p_{[abc]} P_S}.$$

Activation field was calculated within modified model for parameters $R$, $\sigma_{\min}$ and $\delta\sigma$ [31]. Results are shown in the **Figure 4b**. The value of $R$ is chosen in a reasonable agreement with effective pint charge model of the probe [43, 44]. As one can see, $E_a$ is noticeably anisotropic. The dependence of the activation field on the applied voltage $U$ of a threshold-type; the field rapidly decreases with $U$ increase and becomes zero $U > U_{cr}$, indicating the voltage threshold for instant nucleation. The value of $U_{cr}$ depends on $p_{[abc]}$ as following $U_{cr}^X < U_{cr}^Y \ll U_{cr}^X$, because $p_Z < p_Y < p_X$.

We estimated the ratio of the threshold and activation fields in different directions from expressions (4)-(5) for known distances $p_{[abc]}$ and the energy profile of periodic lattice potential. Results are shown in the **Figure 4c and 4d.** The ratio of the threshold fields $E_{th}^X/E_{th}^Y$ is within the range from 1.5 to 5, the ratios $E_{th}^Y/E_{th}^Z$ and $E_{th}^X/E_{th}^Z$ are within the range 1.5 to 100 for actual range of domain wall width. The ratio of the activation fields $E_a^X/E_a^Y$ is within the range from 1.5 to 10, the ratios $E_a^Y/E_a^Z$ and $E_a^X/E_a^Z$ are within the range from 0 to 3.

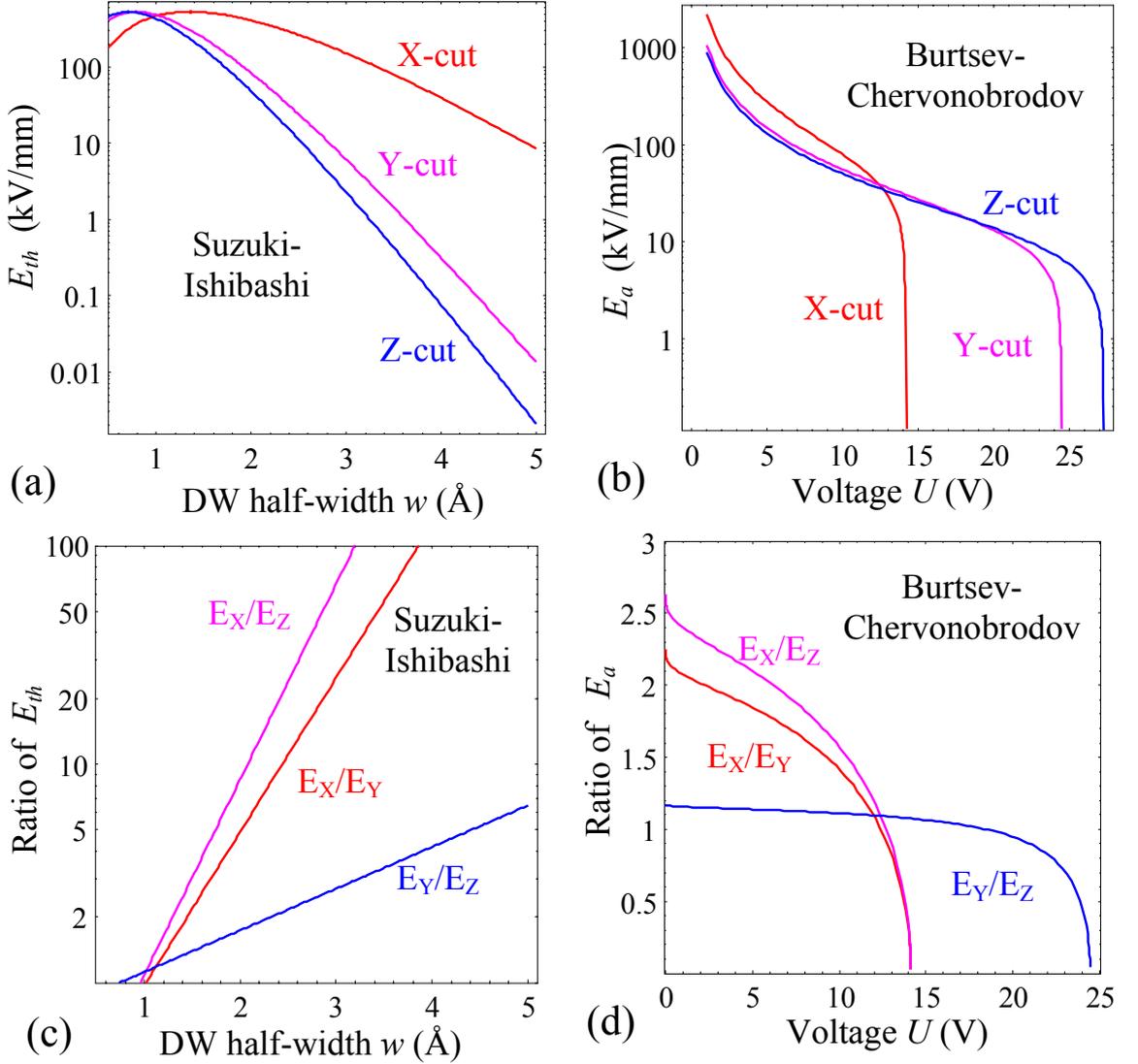

**Figure 4.** (a) Threshold field dependences on the domain wall half-width $w$ calculated within Suzuki-Ishibashi model for LiNbO$_3$ parameters $\alpha = -1.95 \times 10^9$ m/F, $P_S = 0.735$ C/m$^2$. (b) Activation field dependences on the voltage $U$ applied to the probe calculated within Burtsev-Chervonobrodov model for the parameters $R=100$ nm, $\sigma_{min} = 0.160$ J/m$^2$ and $\delta\sigma = 0.150$ J/m$^2$ [31]. (c) The threshold fields' ratio vs. the domain wall half-width $w$. (d) Activation fields' ratio vs. applied voltage $U$.

## 2.3. Theoretical background for interpolation functions of the domain sizes

In accordance with available experiments [7, 8] and nucleation rate theory [24, 45, 46], domain sizes $s(t)$ obey the logarithmic law with the writing time increase, e.g. $s(t) \sim \log(t/t_c)$. Allowing for the existence of critical activation voltage in accordance with B-C model, domain sizes should change rapidly at small writing times, since the domain wall velocity exponentially depends on the dragging electric field [46]. These facts motivate us to intepolate the numerical data by the function $s(t) \sim C^S f(t/t_0) \log(t/t_c - 1) + B^S$ with the fitting parameters $C^S$, $t_0$, $t_c$ and $B^S$. The function $f(t)$ should be transformed into the unity at $t/t_c \gg 1$; while its behavior at small times can be interpolated from the fitting to numerical results. The interpolation functions have sense at writing times $t/t_c > 1$, for which all sizes are positive, indicating e.g. the impossibility to write a stable domain by short pulses.

Asymptotic expression for the angle of the flat domain wall instability is $\theta_f = \arctan\sqrt{\varepsilon_{zz}/\varepsilon_{xx}}$ [47] that gives 19.65 deg for LiNbO$_3$. Note that the expression does not account for the domain wall thickness increase near the charged domain apex and thus it appears essentially higher than the values calculated numerically. Hence, to interpolate the numerical results in dynamics we will use the expression $\theta(t) = C^\theta + B^\theta \exp(-t/t_0)$ for the angle near the domain wall apex, where the value of the constant $C^\theta$ should not coincide with the $\theta_f$ value.

## III. Modeling and interpolation of polarization dynamics

Below we present results for the model case of domain formation under the absence of free carriers; the situation is typical for congruent LiNbO$_3$ without impurities. Appeared that the carriers mostly affect on the domain depth and domain wall conductivity in the ferroelectric-semiconductors [48]. In contrast, the surface shape and sizes evolution is relatively weakly affected by the bulk screening for the carrier concentration less than 10$^{-14}$ cm$^{-3}$. To describe the congruent LiNbO$_3$ ferroelectric and dielectric properties at room temperature we used the following material parameters $\varepsilon_{33}^b = 5$, $\varepsilon_{11} = 84$, $\varepsilon_{33} = 30$, $\alpha = -1.95 \times 10^9$ m/F, $g \sim 10^{-10}$ V·m$^3$/C. Spontaneous polarization $P_S = 0.75$ C/m$^2$ and correlation radius $R_c = \sqrt{-g/\alpha} \approx (0.4 - 1)$ nm. Threshold field vary in the range $E_{th} = (21 - 550)$ kV/mm. The evolution of domain shape and corresponding depolarization field were calculated in COMSOL. Typical results are shown in the **Figure 5.**

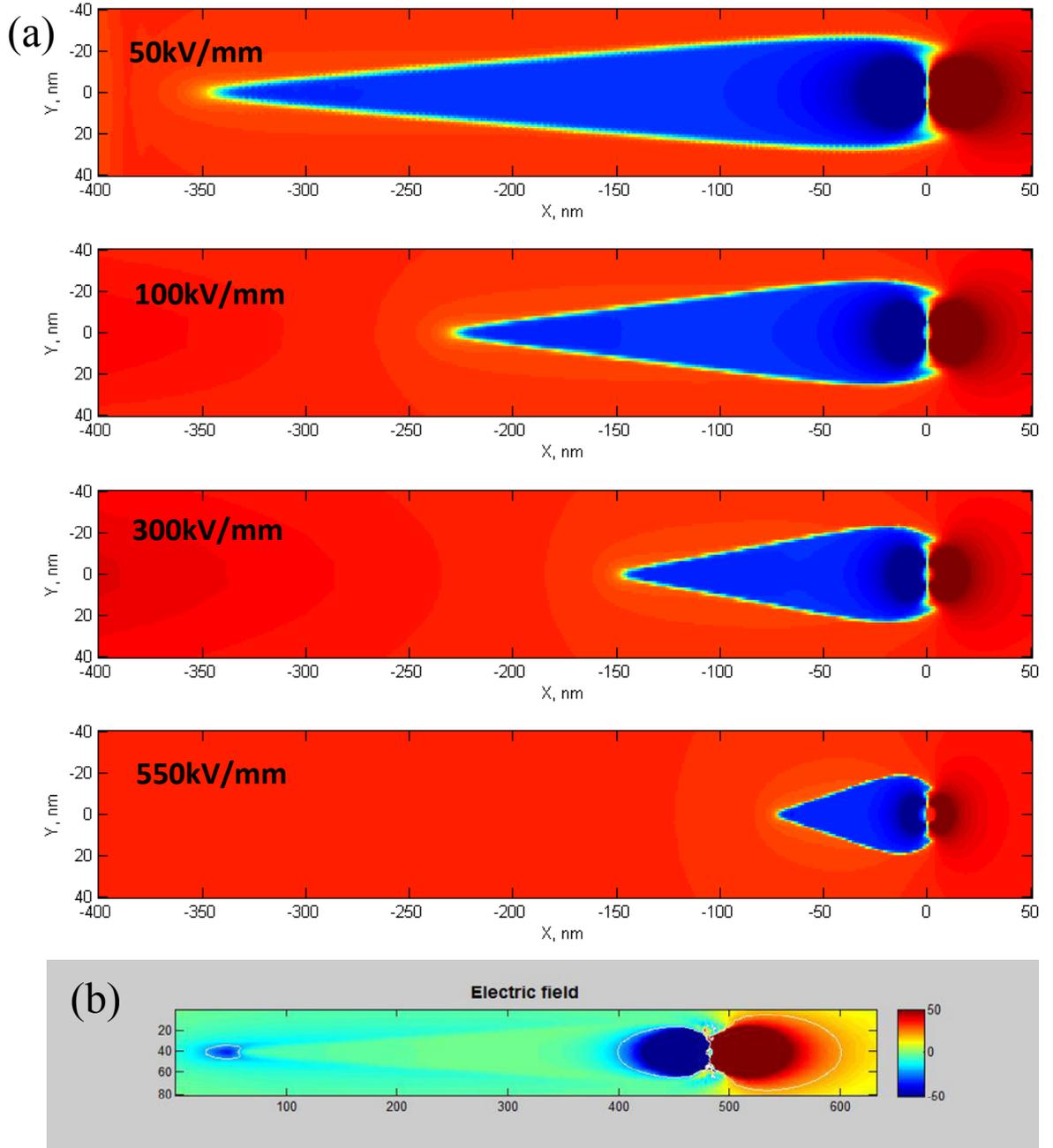

**Figure 5. (a)** Domain shape (top view) calculated for different threshold field, $E_{th}$ = (50 – 550) kV/mm. **(b)** Corresponding depolarization field [32].

The domain shape on the Y-cut (as well as on the X-cut) is close to the cone prolonged in the polar direction Z, at that the length rapidly increases with the threshold field decrease (**Figure 5a**). The shape strongly deviates from the semi-ellipsoidal one, in contrast to the suggestions made earlier in order to obtain analytical expressions for the depolarization electric field energy [34]. The domain wall thickness increases in the immediate vicinity of the charged domain apex in order to decrease the depolarization field that is maximal in the region. Corresponding depolarization field cross-section on the surface is shown in the **Figure 5b**.

Numerical calculations of the electric filed near charged domain wall of the growing domain shown in the **Figure 5b** confirms the breakdown effect and explains domain growth in the areas with external electric field well below the coercive one. Actually the electric field in the immediate vicinity of the domain apex reverses the ferroelectric polarization outside the apex, since it is negative here and higher than the coercive one, while its is positive inside the domain apex. The probe field vanishes in the apex region once the domain lengths exceeds several probe sizes, and the domain apex in principle can move ahead reaching sub-micron and even micron distances until reaching the sample boundaries ("domain breakdown").

In order to perform comparative analyses of the domain sizes and shape evolution, we extracted the temporal dependences of the sizes from the domain profiles calculated in COMSOL. Using designations from the **Figure 1a** we calculated the temporal evolution of the domain apex angle $\theta(t)$; and sizes, with a special attention to the maximal width $r(t)$ and length $l(t)$ on the sample surface, since the surface sizes can be compared with experiment of Alikin et al [33].

Temporal dependencies of the domain length $l$ and width $r$ were calculated in COMSOL for different threshold fields $E_{th}$. Results are shown by symbols in the **Figures 6a-b**. Different color of the symbols corresponds to the different threshold fields $E_{th}$ = (21, 50, 100, 200, 550) kV/mm.

**Figure 6a-b** illustrates the dependences of the domain length and maximal width on the pulse duration in a log-linear scale. The domain sizes monotonically decreases with the threshold field increase. During the activation stage corresponding to the times from 0 to 0.5 $t/\tau_0$ the domain length increases super-linearly, the width increases sub-linearly. Starting from the times $t$>0.5$\tau_0$ all the sizes asymptotically obey the logarithmic law, $l(t) \sim \log(t/t_c)$.

To establish an analytical dependence of the domain sizes vs. writing time we performed the fitting of COMSOL results by using the following interpolation function for the domain length $l$ and width $r$ on the sample surface

$$l(t) = \frac{C_k^l (t/t_{0k})^{3/2}}{(t/t_{0k})^{3/2} + 1} \log((t/t_{ck}) - 1) + B_k^l, \tag{6}$$

$$r(t) = C_k^r \log((t/t_{ck}) - 1) + B_k^r. \tag{7}$$

The subscript $k$=1 – 5 corresponds to the different threshold fields $E_{th}$ = (21, 50, 100, 200, 550) kV/mm. The interpolation functions for the domain sizes have sense for writing times $t/t_{ck} > 1$, indicating the impossibility to write a stable domain by shorter pulses.

**Figures 6c-d** show the dependences of the constants $B_k$, $C_k$ and $t_{ck}$ on the threshold field $E_{th}$. The values of $B_k^{l,r}$ monotonically decrease and $t_{ck}$ monotonically increase with $E_{th}$ increase. As anticipated, the critical times values $t_{c1}$=0.065, $t_{c2}$=0.06, $t_{c3}$=0.05, $t_{c4}$=0.03 and $t_{c5}$=0.02 are the same for the domain length, width and height. The values of $C_k^l$ monotonically decrease, while the values of $C_k^r$ very slightly increases with $E_{th}$ increase.

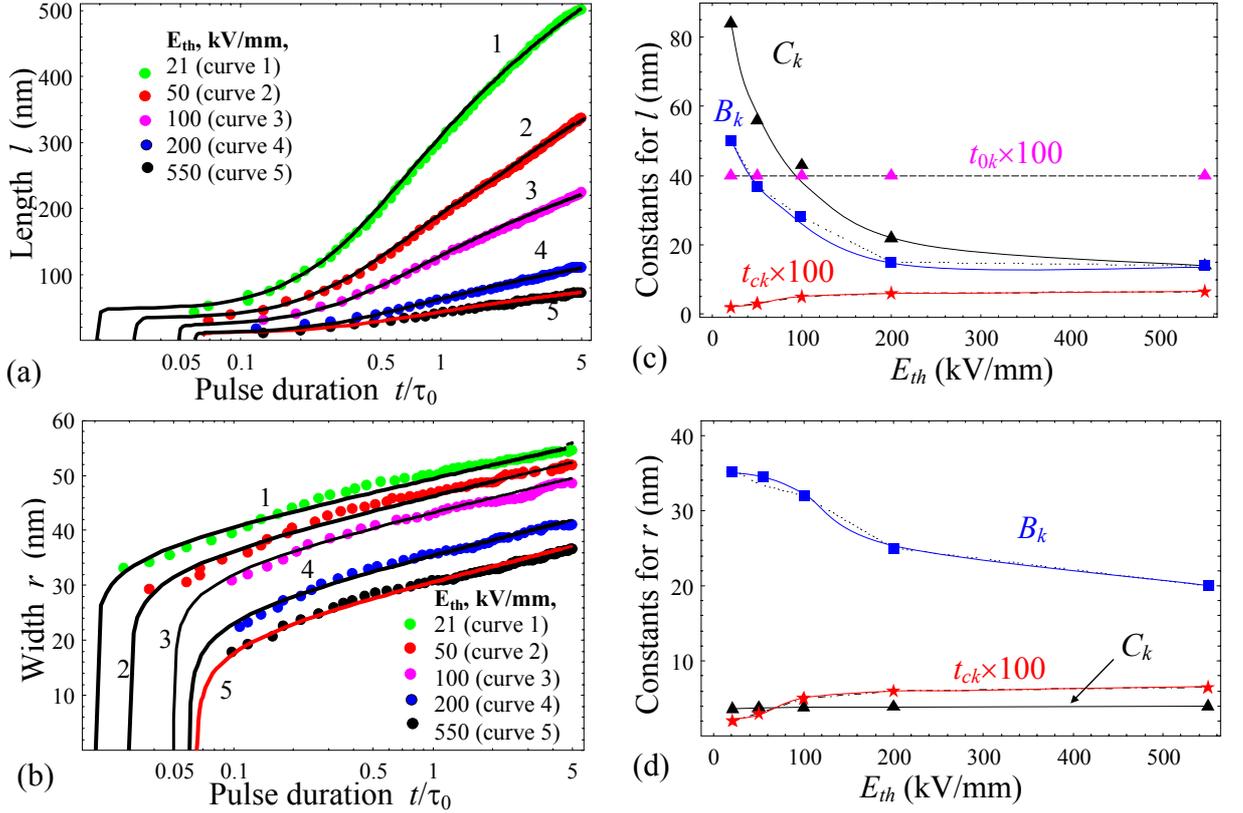

**Figure 6.** Temporal dependencies of the surface domain length **(a)** and width **(b)** calculated for different threshold voltages $E_{th}$ = (21, 50, 100, 200, 550) kV/mm. Points correspond to the numerical results simulated in COMSOL. Solid curves 1-5 correspond to the interpolation functions: **(a)** $C_k^l (t/t_{0k})^{3/2} ((t/t_{0k})^{3/2}+1)^{-1} \log((t/t_{ck})-1) + B_k^l$, plotted for parameters $C_k$, $B_k$, $t_{0k}$ and $t_{ck}$ shown in the plot (c). **(b)** $C_k^r \log((t/t_{ck})-1) + B_k^r$, plotted for parameters $C_k$, $B_k$ and $t_{ck}$ shown in the plot d. The scale for $t_{ck}$ and $t_{0k}$ are $10^2$. **(c,d)** Dependence of the fitting constants $C_k$, $B_k$ and $t_{ck}$ on the threshold field. As anticipated, the critical times $t_{c1}$=0.065, $t_{c2}$=0.06, $t_{c3}$=0.05, $t_{c4}$=0.03 and $t_{c5}$=0.02 are the same for the plots **(a)-(b)**.

Temporal dependence of the angle θ on the dimensionless pulse duration time $t/\tau_0$ was calculated in COMSOL for different threshold voltages $E_{th}$ [**Figures 7**]. To establish the analytical law of the angle dependence on time, we performed the fitting of COMSOL results by the following interpolation function:

$$\theta(t) = C_k^\theta + B_k^\theta \exp(-t/t_{0k}). \tag{8}$$

The subscript $k=1-5$ corresponds to the different threshold fields $E_{th} = (21, 50, 100, 200, 550)$ kV/mm. As one can see from the **Figure 7a** the angle is acute, monotonically decreases and then saturates with time increase in agreement with Eq.(8). As it follows from the **Figure 7b** the constant $B_k$ monotonically increases with the threshold field and then saturates at $E_{th} > 200$ kV/mm. The constant $C_k$ monotonically increases with $E_{th}$ increase; possible saturation can start only at $E_{th} > 500$ kV/mm.

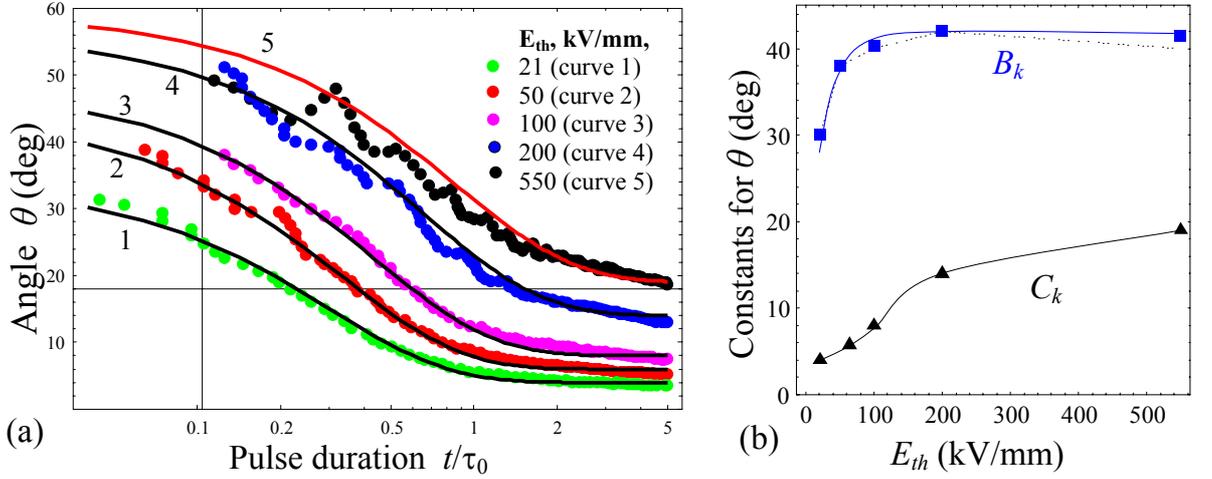

**Figure 7. (a)** Temporal dependence of the domain apex angle θ calculated for different threshold field $E_{th} = (21, 50, 100, 200, 550)$ kV/mm. Points correspond to the numerical results simulated in COMSOL. Solid curves 1-5 are the interpolation functions $C_k + B_k \exp(-t/t_{0k})$ plotted for parameters $C_1=4$; $C_2=6$; $C_3=8$; $C_4=14$; $C_5=19$; $B_1=30$; $B_2=38$; $B_3=40$; $B_4=42$; $B_5=42$; $t_{01}=0.3$; $t_{02}=0.33$; $t_{03}=0.43$; $t_{04}=0.60$; $t_{05}=0.8$. **(b)** Dependence of the constants $C_k$ and $B_k$ on the threshold field.

### IV. Comparison of domain shape and sizes with experiment

Alikin et al [33] measured experimentally the domain shape and sizes on the non-polar X- and Y-cuts. Corresponding domain length and width at the non-polar surfaces of the CLN are shown by symbols with error bars in the **Figures 8a** and **8b**. Solid curves are interpolated functions for the domain sizes given by Eq.(6) and (7) with the best fitting parameters listed in the capture. Using the parameters for domain length we calculated the ratios $B_Y/C_Y = 0.73$ for Y-cut and $B_X/C_X = 0.89$ for X-cut. Regarding the ratios tip radius independent, we can compare the values with the ratio $B/C$ extracted from numerical modeling as it shown in the **Figure 8c-d**. After placing the points $B_Y/C_Y$ and $B_X/C_X$ in the **Fig. 8c-d** we concluded that the threshold field $E_{th}$ for X-cut is about 420 kV/mm and about 250 kV/mm for Y-cut. Note that the best fitting parameters for domain width corresponding to the same values of $E_{th}^X$ and $E_{th}^Y$ is $C_Y=23$, $B_Y=138$, $C_X=15$, $B_X=82.5$. Due to the scattering of the width for X- and especially Y-cut we regard the data for width less reliable than the data for length.

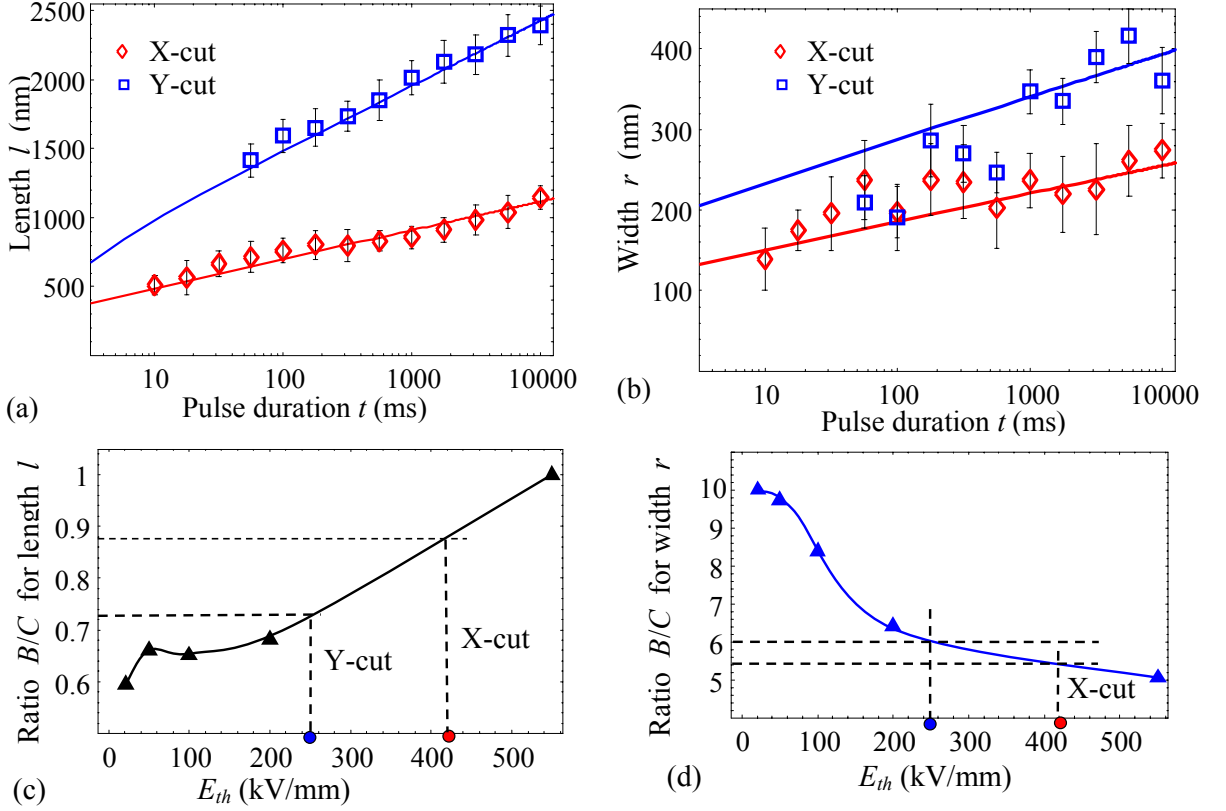

**Figure 8.** Dependencies of **(a)** domain length and **(b)** domain width vs. the switching pulse duration on X- and Y-cuts of LiNbO$_3$. Symbols with error bars are experimental data [33] for X- and Y- cuts of 20-µm-thick CLN placed in dry nitrogen, solid curves are our fitting using the following interpolation functions. The function for the domain length is $\frac{C_k^l (t/t_{0k})^{3/2}}{(t/t_{0k})^{3/2}+1} \log((t/t_{ck})-1) + B_k^l$ with parameters $C_X=90$, $B_X=80$, $t_{cX}=0.1$ ms, $t_{0X}=1$ ms for X-cut and $C_Y=205$, $B_Y=150$, $t_{cY}=0.15$ ms, $t_{0Y}=1$ ms for Y-cut The function for the domain width is $C_k^r \log((t/t_{ck})-1) + B_k^r$ with parameters $C_Y=23$, $B_Y=138$, $t_{cY}=0.15$ ms for Y-cut, $C_X=15$, $B_X=82.5$ and $t_{cX}=0.1$ ms for X-cut. **(c-d)** The ratio $B/C$ extracted from numerical modeling.

Finally, let us discuss the question about the difference in domain depth for the cases when the writing electric field of the probe acts on different polar cuts of LiNbO$_3$. As it was reported earlier by Molotskii et al [8] for the case of nanodomain formation on Z-cut their depth (typically called length because of the radial symmetry of domain cross-section) can reach micron distances due to the breakdown effect. Alikin et al [33] concluded from a selective etching that the domain depth on the Y-cut is rather small in comparison with the one on the X-cut. Moreover, "Y-cut domains" most likely remained nanosized in Y-direction, while "X-cut domain" can be much deeper, but not needle-like as "Z-cut domains".

Note, that the proposed approach accounting for the anisotropy of lattice barriers (with corresponding minimal distance $p_Z \approx 2.310$ Å, $p_Y \approx 2.575$ Å, $p_X \approx 4.469$ Å) and depolarization effects at the charged domain walls can explain these facts. In particular, the longest needle-like shape of Z-cut domains is conditioned by the smallest threshold field $E_{th}(p_Z)$ and domain breakdown in Z-direction. The smallest depth Y-cut domain in X-direction originated from inequality $E_{th}(p_X) \gg E_{th}(p_Y) > E_{th}(p_Z)$, since the smaller is the threshold field the bigger is the domain size. These speculations can be quantified by the $E_{th}(p_{[abc]})$ ratios for different crystallographic cuts shown in the **Figure 4c.**

## V. Conclusion

We explained the physical nature and described quantitatively the anisotropic nanodomain formation induced by a charged the AFM probe on the non-polar cuts of LiNbO$_3$ in the framework of a self-consistent semi-phenomenological semi-microscopic approach. Our theoretical approach takes into account that the height of lattice pinning barriers and the minimal distance between the equilibrium positions of uncharged domain wall should vary in different directions due to the crystallographic anisotropy. In result corresponding threshold field of the domain wall motion is various in different directions. The analysis of atomic positions at polar and non-polar cuts and anisotropy of lattice barriers leads to the conclusion that the threshold field in the YZ-plane (X-cut) should be significantly higher than the one in the XZ-plane (Y-cut) and XY-plane (Z-cut). Corresponding analytical expression for the anisotropic threshold field was obtained within modified Suzuki-Ishibashi approach.

We utilize the relaxation-type differential equation with cubic nonlinearity for the calculation of the polarization dynamics under the probe-induced nanodomain formation. The domain shape in the ZX (or YZ) cross-section is close to the cone prolonged in the polar direction Z. The shape strongly deviates from the semi-ellipsoidal one, in contrast to the suggestions made earlier by Pertsev et al [34]. For intermediate and long pulses the domain sizes logarithmically depend on the pulse duration. We obtained that the smaller is the threshold field the larger are the domain sizes, and the fact allows us to explain quantitatively several times difference in nanodomain length on the X- and Y-cuts observed experimentally in LiNbO$_3$ [33].

Obtained results can give insight into the nanoscale anisotropic dynamics of polarization reversal in strongly inhomogeneous electric field created by the charged AFM probe. In particular, we established that the different velocity of domain growth and consequently equilibrium domain sizes on the X- and Y-cuts of LiNbO$_3$ originate from the anisotropy of the threshold field.

**Acknowledgements**

E.A.E. and A.N.M. acknowledge National Academy of Sciences of Ukraine (grants 35-02-15). A.I. and S.V.K. acknowledge Office of Basic Energy Sciences, U.S. Department of Energy.

**Authors contribution.** A.N.M. formulated and elaborated the theoretical model, derived corresponding analytical expressions and wrote the initial text of the manuscript with illustrations. A.I. performed corresponding numerical simulations of domain shape and sizes in COMSOL, the numerical results processing and smoothing jointly with O.V.V. Y.F. performed fitting of the COMSOL results by interpolation functions. E.A.E. densely worked on the improvement of the anisotropic threshold field model and correlation with experiment of the obtained theoretical results. A.N.M., V.V.O, V.Ya.S., S.V.K. and E.A.E. tightly worked on the physical interpretation of the obtained theoretical results, improvement of the paper text, discussion and conclusions.